\documentclass[twocolumn,showpacs,aps,amsmath,amssymb,pre,superscriptaddress,nofootinbib,10pt]{revtex4}

\usepackage{standalone}
\usepackage[utf8]{inputenc}
\usepackage{soul}
\usepackage{graphicx}
\usepackage{color}
\usepackage{comment}
\usepackage{ulem}

\begin{document}
\title{On the Degeneracy of Spin Ice Graphs, and Its Estimate via  the Bethe Permanent}

\author{Francesco Caravelli}
\affiliation{Theoretical Division (T4),
Los Alamos National Laboratory, Los Alamos, New Mexico 87545, USA}
\author{Michael Saccone}
\affiliation{Center for Nonlinear Studies,
Los Alamos National Laboratory, Los Alamos, New Mexico 87545, USA}
\affiliation{Theoretical Division (T4),
Los Alamos National Laboratory, Los Alamos, New Mexico 87545, USA}
\author{Cristiano Nisoli}
\affiliation{Theoretical Division (T4),
Los Alamos National Laboratory, Los Alamos, New Mexico 87545, USA}

\begin{abstract}
The concept of spin ice can be extended to a general graph. We study the degeneracy of spin ice graph on arbitrary interaction structures via  graph theory. Via the mapping of spin ices to the Ising model, we clarify whether the inverse mapping is possible via a modified Krausz construction. From the gauge freedom of frustrated Ising systems, we derive exact, general results about frustration and degeneracy. We demonstrate for the first time that every spin ice graph, with the exception of the 1D Ising model, is degenerate. We then study  how degeneracy scales in size, using the mapping between Eulerian trails and spin ice manifolds, and a permanental identity for the number of Eulerian orientations. We show that the Bethe permanent technique  provides both an estimate and a lower bound to the frustration of spin ices on arbitrary graphs of even degree. While such technique can be used also to obtain an upper bound, we find that in all the examples we studied but one, another upper bound based on Schrijver inequality is tighter.
\end{abstract}

\maketitle

\section{Introduction}
Ever since the discovery of degeneracy of ground states in constrained, disordered systems obeying the so-called ice rule \cite{bernal,pauling,lieb} and their subsequent experimental implementation in magnetic systems called Spin Ices \cite{ramirez}, there has been an active interest in ice-like frustrated materials. 
Recently, the idea has been extended both theoretically and experimentally to artificial realizations called artificial spin ices (ASI) \cite{colloq} which have allowed for design of frustration to generate exotic behaviors in their collective physics. ASIs  are  arrays of interacting,  shape-anisotropic nano-islands, each of which can be modeled as a binary Ising spin. They can be characterized directly in real time and real space via a variety of techniques~\cite{Wang1,Sandra,Bader}. 

A spin ice can be described abstractly as a set of binary spins arranged on the edge of a lattice, such that its low energy configuration obeys the ice rule. This rule dictates that for each vertex the absolute difference between spins pointing toward the vertex and spins pointing out of the vertex is zero if the vertex has even coordination, or one if the vertex has odd coordination. 

Recently, there has been an intense investigation both in the physical underpinnings and control of artificial spin ices and their emergent interactions \cite{reddim,Heyderman,Canals1,Nisoli1,Morgan,Budrikis,Branford,Ryzhkin,Moeller,Chern2,Le,Chern3,Gliga,Nisoli4,Gilbert2,Bhat,CaravelliMASI}, with a broad interest in applications, ranging from topological order \cite{Castelnovo1,topor}, memory in materials \cite{Lammert2,GilbertMem}, disordered systems and slow relaxation \cite{Cugliandolo2}, novel resistive switching \cite{memristors,caravelli}, and embedding logic circuits in the magnetic substrate \cite{logic2,logic3,logic4,logic5}. These new materials are highly controllable~\cite{gartside,WangYL2,WangYL,colloq,Nisoli4,vavassori0,vavassori} and can be used to realize novel models via engineering geometric frustration \cite{Nisoli4,Schanilec,Nisoli8,mol}. 
 
 The collective behavior of these artificial structures typically depends on the geometry, which is open to design~\cite{Morrison,Nisoli4} via lithographic printing. As novel lithographic techniques are discovered, the control of the dimensionality of these materials requires new theoretical tools to understand the frustration of non-planar artificial spin ices \cite{Ladak3d,Ladak3d2}. Moreover, it is now possible to embed general spin ice graphs into quantum annealers~\cite{QuantumASI}. For these reasons, this paper is focused on a more theoretical and extensible approach for calculating lower and upper bounds to the degeneracy of the ground state of generic spin ices. However, given the fact that we later use the Ising model mapping, it is worth mentioning that the origin of the interaction in an artificial spin ice between the spins is dipolar, and are not exchange interactions as in natural spin ice.

 Previous work~\cite{Field} set up the concept of spin ice on a general graph. Spin ice concepts are often translatable into the language of graph theory, and vice versa. For instance in the mathematical literature a balanced graph (with even degree nodes) is a directed graph whose indegrees is equal to the outdegrees  \cite{Graph1}. In spin ice language, this corresponds to a configuration of the ice manifold~\cite{Field}. Thus, the problem of finding the Pauling entropy~\cite{pauling} of a spin ice graph is therefore equivalent to the problem of counting the number of balanced digraphs. Moreover, it is one of the many celebrated results by Euler that only graphs which can be balanced via an orientation support an Eulerian trail \cite{Euler1}.
In this sense, many results from graph theory can be borrowed to study frustration in spin ice. %Whilst many results are known for balanced graph, less is known for semi-balanced graphs, e.g. graphs whose node vertices are odd and in which the difference between the indegree and the outdegree can be at most one. 

In this paper we use and generalize some of the known results from graph theory.  We demonstrate for the first time that a general spin ice graph is always degenerate, with the exception of the trivial case: the one-dimensional Ising model. Then, because the scaling of degeneracy with size is fundamental to the notion of Pauling entropy in spin ices, we compute lower and upper bound the ground state degeneracy of several spin ices.
%In the case of square ice or its decimations~ \cite{Morrison} the energy can be well approximated using local couplings associated to the interactions between parallel and perpendicular islands $\epsilon_{\perp}>\epsilon_{||}>0$ \cite{Morrison}. 
%Finally, spin ices  of having monopole like charges in  spin ices \textit{without} a string tension
%.

In the first part of the paper, we use the Line Graph dual representation \cite{graph}, and derive properties for the effective Ising model on arbitrary graphs. In particular, we use the Krausz coarse-graining procedure to show that from the effective Ising model there is a well-known technique to obtain the original spin ice interaction graph.

In the second part of the paper, we focus on estimation techniques. For even-degree graphs there exists a permanental identity which in principle, but not in practice, allows for the evaluation of the degeneracy of the spin ice. However, using the Bethe Permanent and the Schrijver bound, it is possible to obtain lower and upper bounds to geometric frustration for the cases in which the degree of the graph is even. We apply these bounds to the square lattice, the triangular tiling, the cubic lattice, and tournaments.

\section{Graph theory for Spin Ice}
General graph theoretic approaches are common tools in Statistical Physics \cite{Baxter,Fisher,CaravelliMarkopoulou}. 
Previously, one of us has discussed the notion of spin ice on a general undirected graph $\mathcal G$, where a spin configuration may be thought of as a directed graph, considered its Coulomb phase properties, and shown that charge correlations are computable from graph spectral analysis~\cite{Field}. 
We use the same approach here, but unlike the previous approximated study, we derive exact results for the degeneracy of the ice manifold. Later in the paper we will also introduce a way to obtain estimates (which are lower bounds), while here we prefer to keep a general discussion. 

Consider a set of spins $s_j$ lying on the bonds of a graph. We define a spin ice Hamiltonian as
\begin{equation}
    H=J \sum_{v} (\sum_{j\rightarrow v}  \pm s_j)^2,
    \label{eq:spinice000}
\end{equation}
where $v$ are the vertices of the graph $\mathcal G$ and each $s_j$ has an orientation specified by $\pm$. Given a vertex $v$, the \textit{charge} of such vertex is defined as $Q_v=\sum_{j\rightarrow v}  \pm s_j$, and thus the Hamiltonian is such that the minimum of the energy corresponds to minimal absolute value of charge, defined as the difference between vertices pointing in or out, as in ref~\cite{Field}.  

To see the connection between frustrated spins in the system consider $e=|E(\mathcal G)|$ the total number of edges  (ordered pairs of adjacent vertices) of the graph~\cite{graph}, while $n=|V(\mathcal G)|$ is the total number of nodes. Let us first introduce a few graph theoretical constructs in order to fix the notation.
 For a generic and undirected graph $\mathcal G$ % of vertices $V=\{v_1,v_2,\cdots,v_n\}$ be the set of vertices of $\mathcal G$ and $E=\{e_1,e_2,\cdots,e_p\}$ be the set of
consider the (undirected) incidence matrix $B$ of size $n \times e$ with entries $B_{i\beta}$, where $i$ is an integer between $1$ and $n$ on the set of vertices and $\beta$ is an integer between $1$ and $e$ on 
the set of edges, such that:
\begin{equation}
  B_{i\beta} = \left\{
    \begin{array}{rl}
      1 & \text{if the edge $\beta$ contains the vertex $i$} ,\\
      0 & \text{otherwise }.
    \end{array} \right.
\label{pij}
\end{equation}
Let us now consider instead the {\it directed incidence matrix} $B_{i\beta}$, of size $n\times p$ constructed as follows. First,  assign an orientation to each edge of the graph. This can be thought as a possible spin configuration~\cite{Field}. Given such orientation $\mathcal O$, we assign the matrix elements of  $B^{\mathcal O}_{v\beta}$ as 
\begin{equation}
    B_{v\beta}=\begin{cases}
    0 & \text{if the edge }\beta \text{ is not incident to the vertex }v\\
    1 & \text{if the edge, given the orientation $\mathcal O$, enter $v$}\\
    -1 & \text{if the edge, given the orientation $\mathcal O$, leaves $v$}
    \end{cases}
\end{equation}
(we use latin indices for vertices and greek for edges or spins).

Crucially, we can rewrite the Hamiltonian for a generic spin ice as
\begin{equation}
    H=J \sum_{v=1}^n (\sum_{\beta=1}^p  B_{v,\beta} s_\beta)^2=J \sum_{v=1}^n \sum_{\beta,\beta^\prime=1}^ep B_{v,\beta}B_{v,\beta^\prime} s_\beta s_{\beta^\prime}
\end{equation}
Swapping the vertex and spin summation, we write
\begin{equation}
    H=J  \sum_{\beta,\beta^\prime=1}^p  Q_{\beta,\beta^\prime} s_\beta s_{\beta^\prime},
\end{equation}
where $Q_{\beta,\beta^\prime}=\sum_vB_{v,\beta}B_{v,\beta^\prime}\equiv (B^t B)_{\beta,\beta^\prime}$ is symmetric.
Do note the following. Given the directed incidence matrix, we have
\begin{equation}
    Q_{\beta,\beta^\prime}=\begin{cases}
    2 &: \text{if }\beta=\beta^\prime \\
    0 &: \text{if the spins $\beta$ and $\beta^\prime$} \\
     &\text{have no vertex in common}\\
    -1 &: \text{if both spins $\beta$, $\beta'$ }\\  & \text{leave or enter a common vertex $v$}\\
    1 &: \text{if one spin $\beta$ leaves a common vertex} \\
    & \text{$v$ and $\beta^\prime$ enters it, and viceversa}.
    \end{cases}
\end{equation}
 Given the matrix $Q$, we can write  $Q=2 I - A$ and define $A$, which is a directed incidence matrix with support on the line graph $\mathcal L(\mathcal G)$. In order to gain some intuition about the matrix $A$, let us discuss the non-directed case first.

\subsection{Undirected Line Graphs}

We start by defining line graphs, which are graph constructed from an undirected graph $\mathcal G$ and such that the edges of the graph $\mathcal G$ become the vertices of the graph $\mathcal L(\mathcal G)$. The edges (or edge) of the line graph are constructed based on the connectivity of the original graph, as follows \cite{Whitney, Krausz,Harary, Beineke}.\footnote{It is interesting to note that here there is a mismatch between the original literature in graph theory, starting with the original work of Harary \cite{Harary} (1965). The original Line Graph of a digraph did not have any negative values, e.g. if the two edges do not have zero sum, then we assign a value zero.}

 Let  $\mathcal G=(V,E)$ denote a graph with vertex set $V=\{v_1,v_2,...v_n\}$ and edge set  $E=\{e_1,e_2,...,e_p\}$ 

\begin{figure}
    \centering
    \includegraphics[scale=1.6]{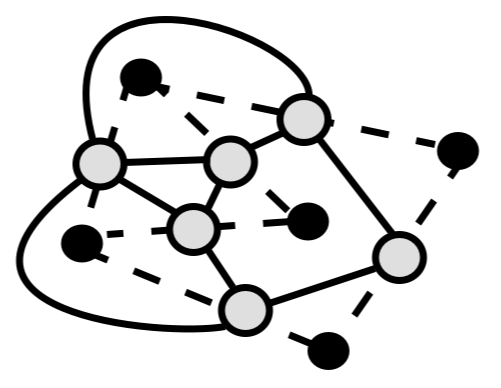}
    \caption{The Line Graph construction. Black vertices and dashed lines correspond the original graph $\mathcal G$, while grey vertices and the solid lines correspond to $\mathcal L(\mathcal G)$.}
    \label{fig:lg}
\end{figure}

Each vertex $\tilde v \in {\widetilde V}({\mathcal L}(\mathcal G))$ corresponds to an edge $e \in E(\mathcal G)$. Two vertices $\tilde v_1$ and $\tilde v_2$ in $\widetilde V(\mathcal L(\mathcal G))$ are adjacent if and only if the edges in $\mathcal G$ (corresponding to $\tilde e_1$ and $\tilde e_2$) share a vertex. The correspondence between $\mathcal G$ and $\mathcal L(\mathcal G)$ is injective but not surjective. From a given graph $\mathcal G$ we can construct only one $\mathcal L(\mathcal G)$, and an example is provided in Fig. \ref{fig:lg}. Yet, in general it is not true that 
any graph can be thought as the line graph or another graph. In fact, according to the Beineke classification, there are 9 non-minimal graphs that are not line graphs of another graph, and each graph containing them is thus not a line graph descendent of any other graph.
We will discuss this later in detail \cite{Beineke}.

Given a graph $\mathcal G$, we can construct its line graph using the following procedure (consider Fig. \ref{fig:lg} for reference):
\begin{enumerate}
\item  Enumerate the vertices of $\mathcal G$. In  Fig. \ref{fig:lg} , these are the black vertices.
\item Enumerate the edges of $\mathcal G$ with a fixed prescription In Fig. \ref{fig:lg} these are the gray nodes, which are the vertices of the line graph $\mathcal L(\mathcal G).$.  
\item If two edges share a vertex, draw a line between them. In Fig. \ref{fig:lg}, these are the solid lines between grey vertices. These becomes the edges of line graph $\mathcal L(\mathcal G).$
\item Remove $\mathcal G$ (nodes and edges) and their enumeration. What is left is the line graph of $\mathcal G$, $\mathcal L(\mathcal G)$. The line graph $\mathcal L(G)$ is thus the set of vertices and edges corresponding to grey vertices and solid lines in Fig. \ref{fig:lg}. 
\end{enumerate}
 %\cn{(COMMENT: do we need to use "blobs"? It reminds me of a programme on RAI3 I used to watch 20 years ago. Pretty good.)}. We could use boob

Consider now the {\it Kirchhoff matrix}, obtained from the (undirected) incidence matrix $B$ of the graph $\mathcal L(\mathcal G).$

%\cn{(COMMENT: we assume above that the reader knows what an incidence matrix, but then we explain it here... Much of the content of this page needs either to be removed or else moved up before these mathematical objects are used.)}
%Later we will generalize this matrix to directed graphs, using the same notation.
The Kirchhoff matrix $K$ is the $p\times p$ matrix built from $B$, such that:
\begin{equation}
K=B^t B,
\label{kirchmatr}
\end{equation}
$B^t$ being the transpose of $B$.
A well-known theorem now gives the relationship between the incidence matrix and the adjacency matrix of the line graph $\mathcal L(\mathcal G)$:

{ {Let $\mathcal G$ be a graph with $p$ edges and $n$ vertices and let $\mathcal L(\mathcal G)$ be its line graph. Then we have}
\begin{equation}
K = A - 2\ I,
\label{eq:kirchhoff}
\end{equation}
where $I$ is the $p\times p$ identity matrix, and $A$ is the adjacency matrix of $\mathcal L(\mathcal G)$}.
\subsection{Directed Line Graphs}
We see immediately that the definition of $Q$ and $K$ are very similar, with an important difference. The matrix $Q$ can be written as
\begin{equation}
    Q=2 I - A,
\end{equation}
where $A$ is called the {\it weighted adjacency matrix}, has the same support as the undirected line graph $\mathcal L(\mathcal G)$, but can take both positive and negative values on edges of the line graph, depending on the orientation $\mathcal O$ (see Fig. 1 and Fig. 2).
In fact, $A$ takes a positive value $+1$ if the two edges have zero sum on the vertex according to the orientation, e.g. if one leaves and one enters, while $+1$ if they both enter and leave.

A fundamental result follows: in general, we can write any spin ice model (written in charge formulation), as
\begin{equation}
    H=-J \sum_{\beta \beta^\prime} A_{\beta \beta^\prime} s_\beta s_{\beta^\prime}
    \label{eq:lg}
\end{equation}
where $A$ is the weighted adjacency matrix according to the rule above. Thus, in the case of directed graphs we can have both ferromagnetic $(J  A_{\beta \beta^\prime}>0)$ and antiferromagnetic values $(J A_{\beta \beta^\prime}<0)$, and is thus a weighted adjacency matrix. If the element of $A$ is positive, the interaction on the line graph is ferromagnetic (e.g. the two spins are aligned in the ground state), while if it is negative the interaction is antiferromagnetic (anti-aligned in the ground state).

Note that spin ices are generally frustrated, but frustration cannot be reabsorbed by a spin redefinition, as it is invariant under the Ising model gauge freedom \cite{Hey} which in our graph-theoretical language corresponds to $s_\beta\rightarrow \xi_\beta s_\beta$, $B_{v\beta} \to  B_{v\beta}  \xi_\beta$ for $\xi_{\beta}=\pm 1$.
As such, the couplings that one obtains in the procedure depend on the gauge transformation. What does not change is the frustration, which cannot be removed. 
As an example, consider Fig. \ref{fig:SquareIce}. Given a certain spin orientation, a 4-vertex node generates an interaction vertex in the equivalent Ising model, which is topologically equivalent to a tetrahedron. There are 3 fundamental cycle, given by $ABD,BDC$ and $ABC$, which are all frustrated because they all contain an odd number of antiferromagnetic interactions.

\begin{figure}
    \centering
    \includegraphics[scale=2.3]{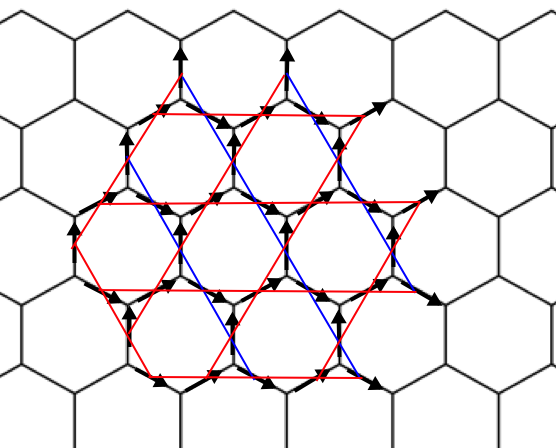}
    \caption{The weighted line graph $A$ superimposed to the Hexagonal lattice, for a given orientation, where blue are antiferromagnetic couplings (negative) and red are ferromagnetic (positive).}
    \label{fig:hexasi}
\end{figure}

\begin{figure}
    \centering
    \includegraphics[scale=1.5]{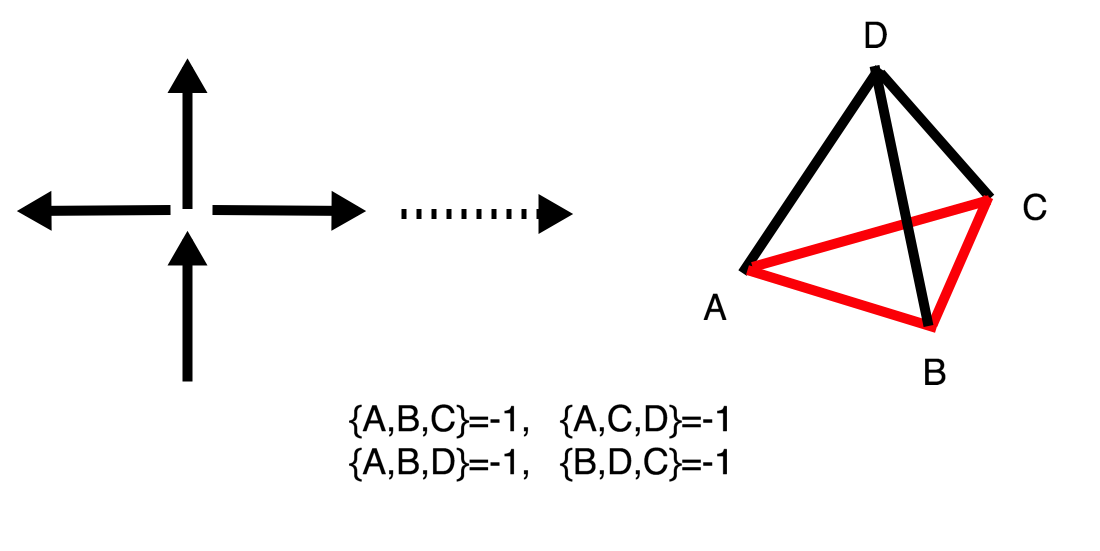}
    \caption{Mapping of a vertex of degree 4 (given a certain orientation of the spins) in the spin ice formulation of eqn. (\ref{eq:spinice000}) to the effective line graph dual of eqn. (\ref{eq:lg}). The figures shows the equivalent Ising interaction model where black lines are ferromagnetic interactions while  with red lines equal to antiferromagnetic interactions. It is possible to see that every single fundamental cycle (the triangles) are frustrated, e.g. the product of the signs of the interactions are always negative.}
    \label{fig:SquareIce}
\end{figure}

Consider now the Hexagonal spin ice of Fig. \ref{fig:hexasi}. We see that with the orientation we have used, the only frustrated cycles are those associated to the vertices of the Hexagonal model. This implies naturally that the degeneracy of the ground state in the model must scale with the size of the vertices.

\subsection{Spin Ice, Frustration, and Degeneracy: General Facts }

We are now in a position to state some general facts for a spin ice on a graph.

\textbf{Remark 1} \textit{Not all frustrated Ising models are spin ices.}
Because the directed line graph dual has support on the undirected line graph, we can borrow the results from the undirected case \cite{Beineke}. If the line graph contains any of the subgraphs contained in the Beineke classification, then we know that the graph is not the line graph of any root graph.
%\begin{figure}
%    \centering
%    \includegraphics[scale=0.1]{Beineke.png}
%    \caption{The nine subgraphs which are not the line dual %of any graph.}
%    \label{fig:beinekeclass}
%\end{figure}

\textbf{Remark 2} \textit{Vertices are mapped to complete graph interactions.}
This is a well known fact that we restate graph-theoretically. Vertices in a spin ice are mapped to a complete graph with a number of vertices equal to the degree of the vertex. This implies immediately that if a spin ice is composed by a sequence of vertices of degree $d=\{d_1,\cdots, d_n\}$, and if for any $i$ we have $d_i>3$, the line graph dual will not be planar because complete graphs on four vertices are necessarily not planar. 

\textbf{Remark 3} \textit{The only purely ferromagnetic spin ice graph is the one dimensional {Ising model. It is also the only spin ice whose ground state is non-degenerate.}}
Consider the following one-dimensional spin ice:
\begin{equation}
    H=J \sum_{i=1}^n (s_i-s_{i+1})^2.
\end{equation}
The ground state has a $Z_2$ symmetry: these are all right or all left spins, which is equivalent to a 1-dimensional ferromagnetic Ising model. This implies that at least one spin ice is non-degenerate. Interestingly, it is the only one. 
%\cn{(Comment: is it custumary to use $D$ for the degree/coordination of a vertex?)}

In order to see this, consider a vertex with $d$ edges (or $d$ incoming/outgoing spins). The total number of interaction terms in the effective Ising model are $d (d-1)/2$. Assume  an orientation in which $d_1$ spins point in and $d_2$  out, with $d=d_1+d_2$. Then, the number of ferromagnetic and antiferromagnetic interactions are, following the directed line graph construction,
\begin{eqnarray}
\text{antiferromagnetic }&:&    \frac{d_1(d_1-1)}{2}+ \frac{d_2(d_2-1)}{2}, \nonumber \\
\text{ferromagnetic} &:& d_1 d_2.
\end{eqnarray}
The only case in which we have no antiferromagnetic interaction is $d_1=d_2=1$, which is a vertex of degree $2$. The only (connected) graphs that can be formed with vertices of  degree two are circle graphs. Thus, one dimensional spin ices are the only purely ferromagnetic models, thus unfrustrated. This implies that all spin ices in dimensions higher than one are necessarily frustrated. 
%% should we explicitly state the sole existence of ferromagnetic interactions implies no frustration?  

This does not mean that all models are necessarily {\it extensively degenerate}, that is possess  a nonzero Pauling entropy \cite{bernal,pauling}. This is a more complicated notion which depends on the product of signs of interaction on a loop. We discuss this next.

\textbf{Remark 4:} \textit{All spin ices with $d>2$ have frustration at the  vertex level}.

Since frustration is gauge invariant, we can pick any orientation of the vertex configuration and calculate frustration along a certain loop (a closed sequence of edges, or a loop) in that particular configuration.
Let us choose $d_2=0$, thus all spins going into the vertex. %As such, all interactions in the vertex are antiferromagnetic. 
Now, all the fundamental cycles at the vertex interactions are of length 3, as the effective interaction is a complete graph $K_d$. There  are $m=\frac{d(d-1)(d-2)}{3!}$ fundamental circuits of length $3$. Since all interactions are anti-ferromagnetic, we have that the product of the signs in every cycle is $-1$, and thus frustrated.

\textbf{Remark 5:} \textit{For planar spin ices, frustration is only at the vertex level.}

This is a byproduct of the following fact. Consider a cycle in a spin ice (in the original lattice). We can always choose an orientation of the lattice such that, for a given cycle, the arrows are chosen head to tail. Thus, when we construct the directed line graphs, the interactions bordering two spins are going to be ferromagnetic. Thus, if we go around a cycle in the line graph dual, we only have ferromagnetic interactions, and thus the cycle is not frustrated. In order to see that this is true always, note that for a planar graph we can always choose orientations of the spin such that such configuration is consistent. Thus, frustration is due only to vertices (cliques) in the directed line graph dual.

Note that this implies that so-called vertex-frustration~\cite{Morrison}, that is the inability to arrange collectively all vertices in a lowest energy configuration, cannot exist in a graph spin ice whose Hamiltonian depends only on the vertex charge. Indeed, all the vertex-frustrated systems \cite{Morrison}, many of which have been realized \cite{tetris,shakti} and depend upon a lifting of degeneracy within vertices of the same charge. They are therefore not pure spin ice graphs.

\subsection{Spin ice reconstruction via Krausz clique partitions}

One of the most interesting byproducts of the direct construction is that there is an inverse procedure, known as Krausz partitioning. We know that if the original spin ice interaction is planar, then vertices are mapped to fully frustrated \textit{cliques} (condition 1). Also, any cycle subgraph which is not a clique must not be frustrated (condition 2). If these conditions are satisfied, and if none of the Beineke graphs are present, then we can reconstruct the original spin ice interaction via the Krausz decomposition, which goes as follows (condition 3). 

\begin{figure}
    \centering
    \includegraphics[scale=1.4]{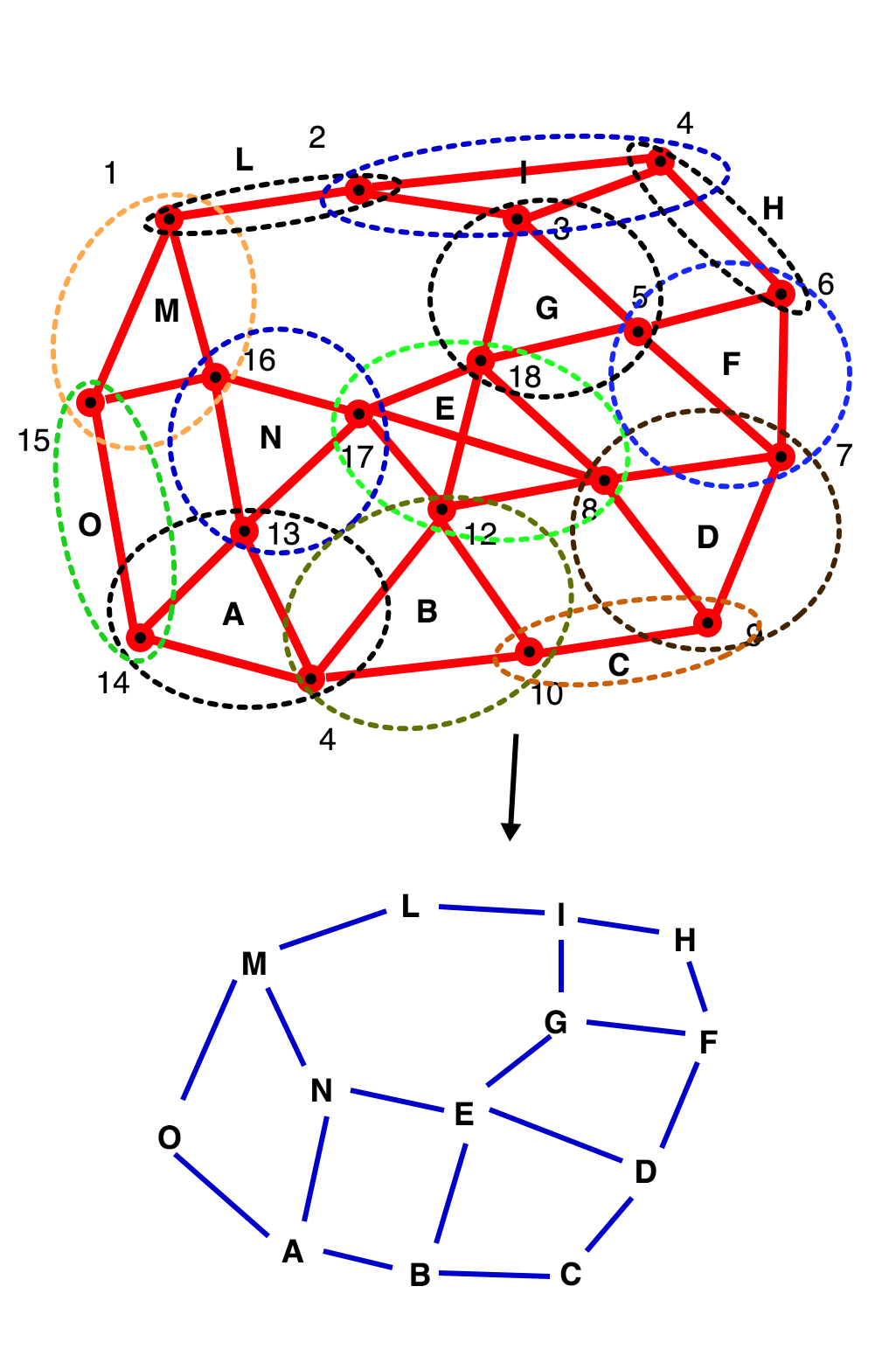}
    \caption{The coarse graining procedure according to the Krausz partitioning.}
    \label{fig:Krausz}
\end{figure}

Given a graph $\mathcal Q$, if condition $1$ and $2$ are satisfied, consider the unweighted graph $|\mathcal Q|$, and
\begin{enumerate}
    \item Enumerate all complete subgraph of the graph $|\mathcal Q|$, and defined as partitions $\mathcal K$;
    \item If all partitions $\mathcal K$ have only one vertex in common, contract the cliques into a vertex, and connect the partitions by an edge
    \item The resulting graph is the spin ice interaction matrix: assign spins to the edges of the resulting graph and add a term $J (\sum_i s_i)^2$ to the corresponding interaction.
    \item Because of gauge invariance, the directionality of the original spin is irrelevant.
\end{enumerate}
It is important to note that a graph is a line graph of a root graph $\mathcal G$ if and only if there is a Krausz partitioning of the graph $\mathcal Q$.
An example of such procedure is in Fig. \ref{fig:Krausz}. Each complete subgraph is identified and the coarse graining procedure is performed. 

\section{Lower and upper bounds to spin ice degeneracy}

Extending Pauling's estimate~\cite{pauling} can become extremely challenging on arbitrary graphs. In order to  estimate the degeneracy of the ice manifold beyond the case of planar graphs or non-bipartite graphs, we will use a general graph-theoretic approach and upper bound the entropy associated to the ice manifold. First, let us note that the maximum degeneracy that a spin ice can have is given by $2^{N_{spins}}$. If the spin ice is degree regular and has $N_v$ vertices, then the maximum entropy of the ice manifold is naturally given by $\epsilon_{max}=N_v \frac{d_v}{2} \ln 2. $
Below we provide a procedure to systematically calculate upper bounds based on the theory of Eulerian tours on graphs. %We  adapt this theory to our benefit in order to deal with graphs that are not necessarily bipartite or planar. 
Given a certain spin ice graph $\mathcal G$, we are interested in calculating the number of configurations in the ice manifold, $\epsilon(\mathcal G)$, and its entropy $S(\mathcal G)=\ln \epsilon(\mathcal G)$ \cite{pauling}.

%\subsection{Case of even degrees}
We focus on the case in which all vertices have even degree, independently from the planarity of the spin ice.
%Before we proceed, let us introduce a few definitions which will be useful later, and that will see to be connected to the problem of frustration in spin ice.
Consider a graph $\mathcal G$. It is very well known that Euler got interested in the problem of walks on graphs with the following property: starting from a certain vertex $v$, perform a walk on the graph $\mathcal G$ such that you never use the same edge twice. A Euler cycle starts at a vertex $v$ and ends in the vertex $v$. Euler proved the following theorem \cite{Euler1}:

\textbf{Theorem} (Euler) \textit{Let $\mathcal G$ be a connected graph. Then, $\mathcal G$  has an Euler cycle if and only if every vertex $v$ has even degree $d_v$}.

\begin{figure}
    \centering
    \includegraphics[scale=1.4]{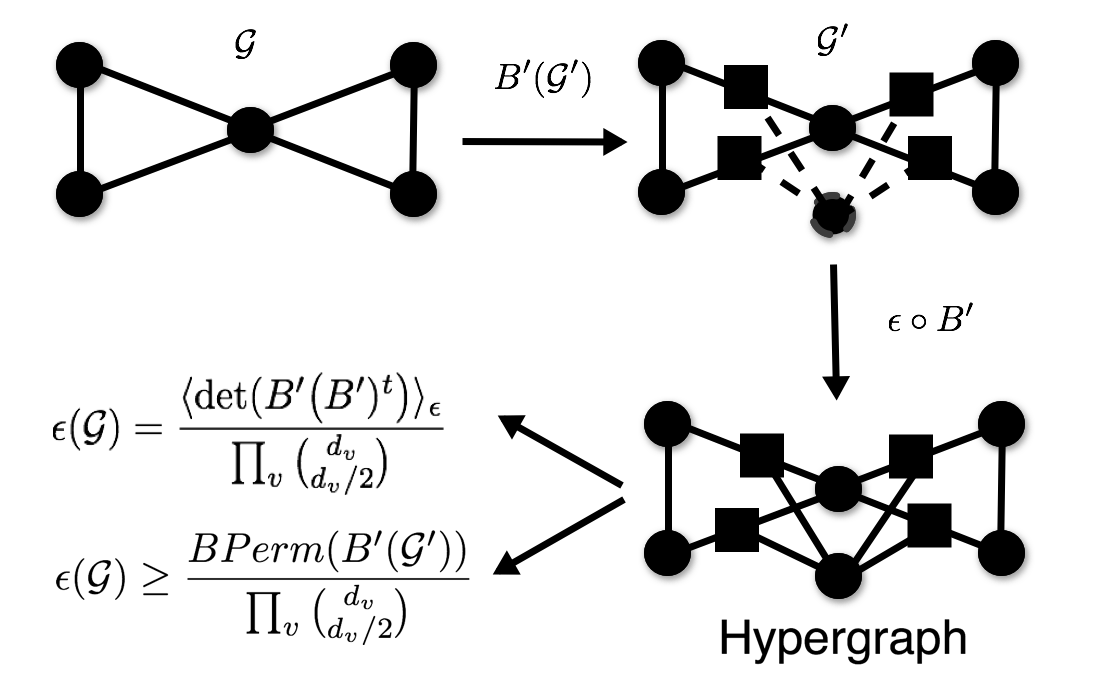}
    \caption{Hypergraph construction for the spin ice degeneracy.}
    \label{fig:deg}
\end{figure}

One direction of this theorem is rather obvious, as if a Euler cycle exists, necessary $d_v$ must be even. The theorem is powerful because it ensures that if all $d_v$ are even, the converse also applies. While an enumeration of the number of eulerian tours for undirected graph is an open problem, a formula for the number of eulerian orientations exists.
Let $\mathcal G$ be a graph with degrees $d_v$. Then, the total number of Eulerian orientations $\epsilon(\mathcal G)$ is given by \cite{schrijver}:
\begin{eqnarray}
\epsilon(\mathcal G)=\frac{\text{perm}(A)}{\prod_{v\in V} (\frac{d_v}{2})!},
\label{eq:perme}
\end{eqnarray}
where the matrix $A$ is the incidence matrix of a hypergraph, whose construction we discuss below. The identity above provides a pathway towards the estimate of the degeneracy of a spin ice graph. The definition of the matrix $A$  is the following (we refer to Fig. \ref{fig:deg}).

Let us consider the undirected incidence matrix $B$ of the undirected graph $\mathcal G$, which represents the spin ice graph and which was introduced in eqn. (\ref{pij}) for a graph $\mathcal G=(V,E)$ which contains $|V|$ vertices and $|E|$ edges. By construction, the incidence matrix $B$ contains $|V|$ rows, one for each vertex of the spin ice, and $|E|$ columns, one for each edge of the spin ice, e.g. the number of spins.  For each vertex $v\in V$ (which is a vertex in the spin ice), $A$ contains a $\frac{d_v}{2}$  number of identical rows of $B$. This implies that $A$ is square and of size equal to the number of edges of $\mathcal G$.
The graphical construction of the matrix $A$, which can be associated to an hypergraph, is shown in Fig. \ref{fig:deg}. The matrix $A$ is thus the incidence matrix of a certain hypergraph $\mathcal G^\prime$, where the number of edges is repeated depending on the degree of the nodes $v$.  For instance, for the graph $\mathcal G$ of Fig. \ref{fig:deg} and the enumeration of the vertices and edges, the incidence matrix $B$ and the matrix $A$ are given by
\begin{equation}
    B=\begin{pmatrix}
    1 & 1 & 0 & 0 & 0 & 0\\
    1 & 0 & 1 & 0 & 0 & 0\\
    0 & 0 & 0 & 1 & 1 & 0 \\
    0 & 0 & 0 & 0 & 1 & 1 \\ 
    0 & 1 & 1  & 1 & 0 & 1 
    \end{pmatrix}\rightarrow A=  \begin{pmatrix}
    1 & 1 & 0 & 0 & 0 & 0\\
    1 & 0 & 1 & 0 & 0 & 0\\
    0 & 0 & 0 & 1 & 1 & 0 \\
    0 & 0 & 0 & 0 & 1 & 1 \\ 
    0 & 1 & 1  & 1 & 0 & 1 \\
    0 & 1 & 1  & 1 & 0 & 1 
    \end{pmatrix}
\end{equation}
where we see that the last row of $B$, associated to the vertex $5$ has been doubled, adding a new node to the graph and making $A$ a square matrix of size $6\times 6$, where $|E|=6$. Thus, the evaluation of the permanent depends on the number of edges of the graph rather than the number of nodes. The addition of the two nodes however implies that the edges are connected to three nodes.  If we force the interpretation of the matrix $A$ as an incidence matrix,  then it is the incidence matrix of an hypergraph $\mathcal G^\prime$, as an effective edge of the system can be connected to more than two nodes. For instance, the edges $2,3,4,6$ are effectively connected to 3 nodes (the edges in hypergraphs can connect to multiple nodes, while only to two in simple graphs).

The formula of eqn. (\ref{eq:perme})  is exact for graphs of even degree but hard to use in practice. This is due to the fact that the permanent is rather hard to calculate numerically for arbitrary graphs, being the problem $\#P$-complete \cite{valiant}. However, one can upper bound the permanent of $(0,1)$ matrices using the Bregman-Minc result \cite{schrijver}, which is given by the following inequality:
\begin{eqnarray}
\text{perm}(A)\leq \prod_{i=1}^m (r_i!)^{\frac{1}{r_i}},
\end{eqnarray}
where $r_i$ is the row sum of the i-th row of $A$. This implies that, if $d_v$ is the degree of the graph, one has Schrijver's inequality  \cite{schrijver}
\begin{eqnarray}
\epsilon(\mathcal G)\leq \prod_{v\in V} \sqrt{\binom{d_v}{\frac{d_v}{2}}}.
\end{eqnarray}

We note that the upper bound above is base on the fact that the graph has an eulerian orientation, and thus $d_v$ must be even. For degree regular graphs, we have
\begin{eqnarray}
\ln \epsilon(\mathcal G)\leq \frac{N}{2} \ln\binom{d_v}{\frac{d_v}{2}}.
\end{eqnarray}
which is a bound of fairly general nature for the number of eulerian paths, and depends on graph ''local" properties, such as the vertex degree.

\subsection{Approximating the number of Eulerian configurations from the permanent}
%Since evaluating the permanent is a $\#P$-complete problem in general, for general graphs one has to resort to some approximations.

%\subsubsection{Numerical estimates}
For completeness, we discuss first a technique which proved unfruitful for us, but which deserves to be mentioned. One way to calculate such quantity exploits the Godsil-Gutman theorem \cite{GGe,Lovasz}. Let $A$ be a non-negative matrix. Then, if we define $B_{ij} =r{ij} \sqrt{a_{ij}}$, where $r{ij}$'s are uncorrelated random variables distributed according to $P(r_{ij})$, with mean $0$ and variance $1$, we have 
\begin{eqnarray}
\text{perm}(A) = \langle \text{det}B^t B \rangle_{P(r)}.
\end{eqnarray}
Since we are interested in the logarithm, we have
\begin{eqnarray}
\ln \text{perm}(A)= \ln \langle \text{det}(B)^2 \rangle_{P(r)}=\ln \langle \text{det}(B^t B) \rangle_{P(r)}. %\geq \langle \ln \text{det}(BB^t)\rangle.
\end{eqnarray}
If $r_{ij}=\{1,-1\}$ the estimator above is called Godsil-Gutman, but if $r{ij}\in \mathcal N(0,1)$ is is called Barvinok estimator.
We have tested both the Gutman-Godsil and Barvinok estimators for the case of the triangular, square and cubic lattice degeneracies, but we have found that the variance of the estimates does not fall fast enough with the number of Monte Carlo samples. 

The second approximation method for the permanent of a non-negative matrix is based on Belief Propagation, which is the one we present here~\cite{huang}. 
Consider the square matrix $A$ obtained via the Schrijver augmentation. The permanent of the matrix $A$ is defined via the sum over all possible permutations, as
\begin{equation}
    \text{perm}(A)=\sum_{\pi\in S_n} f(\pi; A),
\end{equation}
where $f(\pi;W)=\prod_{i=1}^n A_{i \pi(i)}$. Another way of thinking of these permutations is in terms of perfect matching between two sets $A$ and $B$, and in particular ``double dimerizations", as follows. 

Given a certain permutation $\pi$,  a matching between the set A (the first index) and the set B (the second index), can be represented as $\Big(i,\sigma(i)\Big)$. Similarly, a particular valid configuration of the permanent is a set of $n$ non-overlapping dimers
$A_{1,\sigma(1)}....A_{n,\sigma(n)}$. So one constructs a bipartite graph in which one places a dimer between $(i,j)$ if $ A_{ij}>0$. The permanent is a sum over all possible dimerizations.

Based on this idea, Huang and Jebara mapped the permanent to a set of double-dimerizations which correspond to a valid choice of the permanent as follows \cite{huang}. A dimer between the set $A$ and $B$ is an assignment between the variables $X=\{x_1,\cdots,x_n\}$ and the variables $Y=\{ y_1,\cdots,y_n\}$.
We now introduce the potentials $\phi(x_i)=\sqrt{A_{i x_i}}$ and $\phi(y_j)=\sqrt{A_{y_j j}}$ and introduce the function 
\begin{eqnarray}
f(X,Y)=\prod_{ij} \psi(x_i,y_j) \prod_{k} \phi(x_k) \phi(y_k)
\label{eq:factg}
\end{eqnarray}
which is a function of the assignment. If the function $\psi(x_i,y_j)$ enforces that, given two sets of assignments (two possible dimer configurations) between $A$ and $B$, the two assignments are identical, then ones obtains 
\begin{eqnarray}
\prod_{k} \phi(x_k) \phi(y_k)=A_{1,\sigma(1)}....A_{n,\sigma(n)},
\end{eqnarray}
and 
\begin{eqnarray}
Z\equiv\sum_{\sigma,\pi\in S_n} f(X,Y).
\label{eq:partfun}
\end{eqnarray}
where $\sigma=(i,x_i)$ and $\pi=(y_j,j)$. In terms of the dimer representation, a valid configuration is such that two dimers either overlap completely or do not overlap at all. Then, a logic function which ensures such condition is the negation of the XOR function $I(\neg (j= x_i \oplus i=y_j))$, where the function $I(\cdot)$ is zero if the condition is false and  one otherwise.

Note that eqn.~(\ref{eq:partfun}) can be interpreted as the partition function for a particular factor graph with pairwise interactions, and can be analyzed in terms of  belief propagation \cite{Pearl}. The partition function can be interpreted as products of probability distributions which can be decomposed via Bayes' theorem. The key idea behind the BP algorithm is to factor the marginal probability with respect to a certain variable in a product of contributions coming from ``neighboring"
variables; these factorized probabilities are called the messages. One can imagine these factorizations as ``messages" being sent to from neighboring variables $\mathcal N(i)$, informing the full distribution. Borrowing the language used in statistical physics, the variables are called vertices, as the method is applicable exactly to Ising models on trees. The messages sent by a vertex $i$ to
$j\in \mathcal N(i)$ depends on the messages it received previously
from other vertices. Let us assume that the variables $x_i$ take values in a certain collection $F$. Then:
\begin{equation}
m_{i\to j}(x_{j})
\leftarrow
\sum_{x_i\in F} n_{i\rightarrow j}(x_i) 
\phi_i(x_i)\psi_{ij}(x_i, x_j),
\end{equation}
where
\begin{equation}
n_{i\to j}(x_i)=\prod_{k\in \mathcal N(i)\setminus\{j\}}
m_{k\to i}(x_i).
\end{equation}
Messages are positive and satisfy 
\begin{equation}\label{eq:normalization}
 \sum_{x_j\in F} m_{i\rightarrow j}(x_j)=1,
\end{equation}
and thus can be interpreted as probability distribution.
Belief propagation outputs \emph{beliefs}, which are
approximations of the one-site and two-site marginal distributions of $p(\vec x)$.
The beliefs $b_i$ are reconstructed according to
\begin{equation}\label{belief1}
b_i(\tau_i) \propto 
\phi_i(x_i)\prod_{j\in \mathcal N(i)} m_{i\to j}(x_i)
\end{equation}
It has been realized that the fixed
points of the BP algorithm coincide with local minima of the Bethe
free energy) \cite{YeFrWe} .
The Bethe free energy $F_\text{Bethe}$ is given by
\begin{eqnarray}
F_\text{Bethe}&=&-\sum_{ij} \sum_{x_i,y_k} b(x_i,y_j) \ln \Big(\psi(x_i,y_j) \phi(x_i)\phi(y_j)\Big)\nonumber \\
&+&\sum_{ij} \sum_{x_i,y_j} b(x_i,y_j) \ln b(x_i,y_j) \nonumber \\
&-& (n-1) \sum_i \sum_{x_i} b(x_i) \ln b(x_i)\nonumber \\
&-&(n-1) \sum_j \sum_{y_j} b(y_j) \ln b(y_j)
\end{eqnarray}

Thus,  the partition function can be written as a minimization of the Bethe free energy
\begin{eqnarray}
Z=e^{-\text{min}_b F_\text{Bethe}(b)}.
\end{eqnarray}
The minima of the belief can thus be obtained via the message passing algorithm and the beliefs must satisfy $b(x_i)=\sum_{y_j} b(x_i,y_j)$ and $b(y_j)=\sum_{x_i} b(x_i,y_j)$, and $\sum_{x_i,y_j} b(x_i,y_j)=1$, and these functions can be obtained iteratively as
\begin{equation}
    b(x_i,y_j)\propto \psi(x_i,y_j) \phi(x_i)\phi(y_j) \prod_{k\neq j} m_{y_k}(x_i) \prod_{l\neq i}m_{x_l}(y_j)
\end{equation}
and, following the equations described earlier, we have
\begin{eqnarray}
    b(x_i)&\propto&\phi(x_i) \prod_{l\neq i}m_{y_l}(x_i) \nonumber \\
   b(y_j)&\propto& \phi(y_j) \prod_{l\neq i}m_{x_l}(y_j).
\end{eqnarray}
The messages can be obtained iteratively, starting from a random initial state
\begin{eqnarray}
m_{x_i}^{t+1}(y_j)=\sum_{x_i}\Big(\phi(x_i) \psi(x_i,y_j) \prod_{k\neq j} m_{y_k}^t(x_i)\Big).
\end{eqnarray}
On trees, these equations always converge towards the exact solution. Otherwise, the solution is only approximate. This said, an interesting byproduct is that it is possible to prove that the Bethe permanent can be used both
for a lower and an upper bound \cite{vontobel,gurvits}
to the permanent, given the following:

\textbf{Theorem} (Vontobel-Gurvits)
\begin{eqnarray}
\text{\text{Bperm}}(A)\leq \text{perm}(A)\leq \sqrt{2}^n\text{\text{Bperm}}(A).
\end{eqnarray}
where $n$ is the size of $A$ and $\text{Bperm}(A)$ is its Bethe permanent.

The theorem above implies that we can obtain, via the Bethe permanent Bp$(A)$, a level of confidence on the value of the permanent and in particular a certificate for the lower bound scaling.
Note that in practice, we have found that Schrijver's upper bound is typically lower than the one obtained via the Bethe Permanent.

Given a certain lattice described by the graph $\mathcal G$ we call $B\epsilon(\mathcal G)$ the lower bound obtained via the permanent. We thus have
\begin{eqnarray}
B\epsilon(\mathcal G)\leq \epsilon (\mathcal G)\leq \prod_v \sqrt{ \binom{d_v}{\frac{d_v}{2}}}, 
\end{eqnarray}
where $d_v$'s are the vertex degrees.

Using the bounds above, we can then constrain the frustration of the graph via the numerical evaluation of the Bethe permanent. We study the square lattice, the triangular tiling, and the cubic lattice, all with toroidal boundary conditions. The reason is that with these boundary conditions every node has a degree which is even. Thus, boundary effects for small lattices like ours are negligible in this setting. Other boundary conditions are possible, but so far as every vertex degree is even.

\begin{figure}
    \centering
    \includegraphics[scale=0.16]{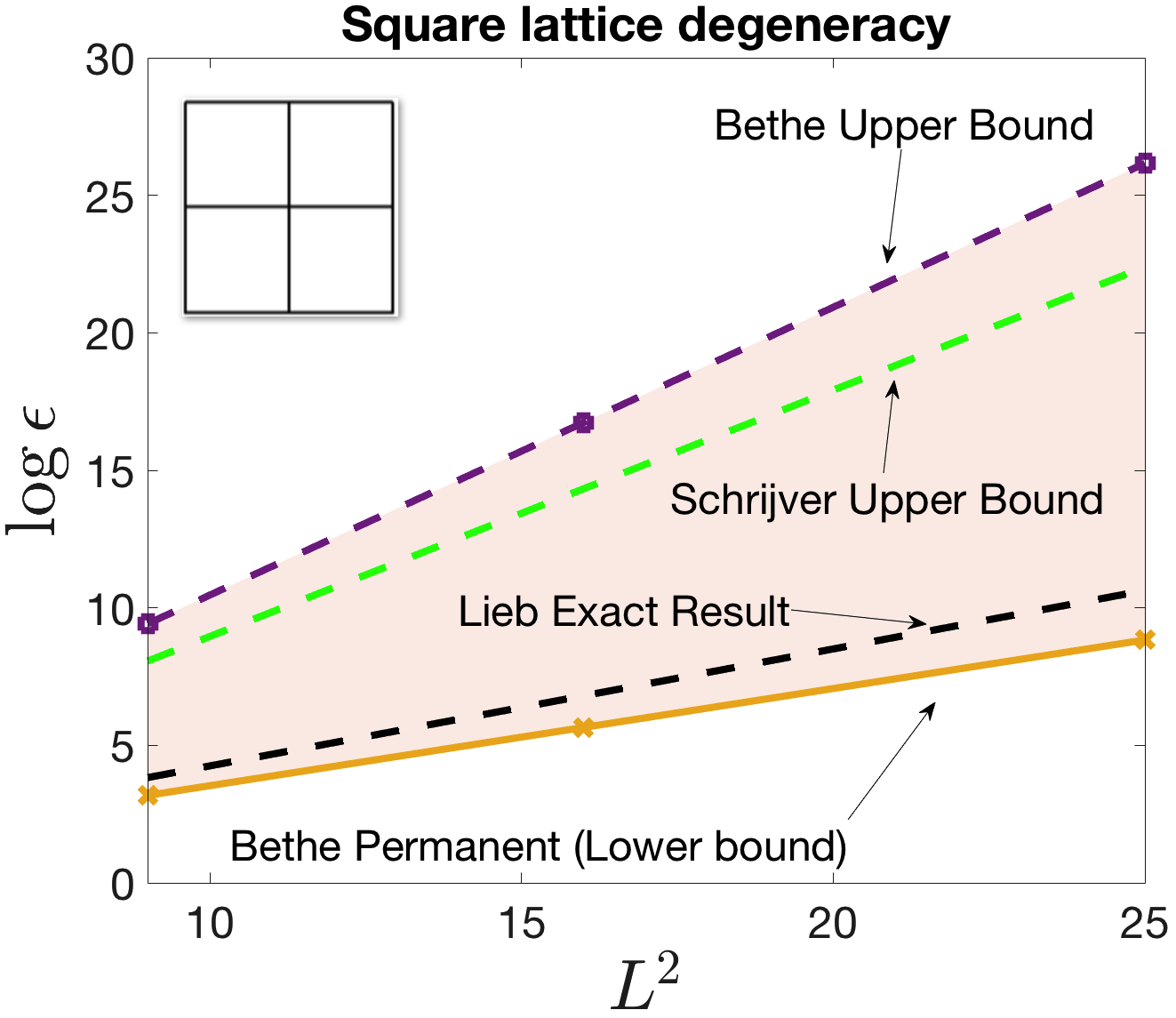}\\
    \includegraphics[scale=0.16]{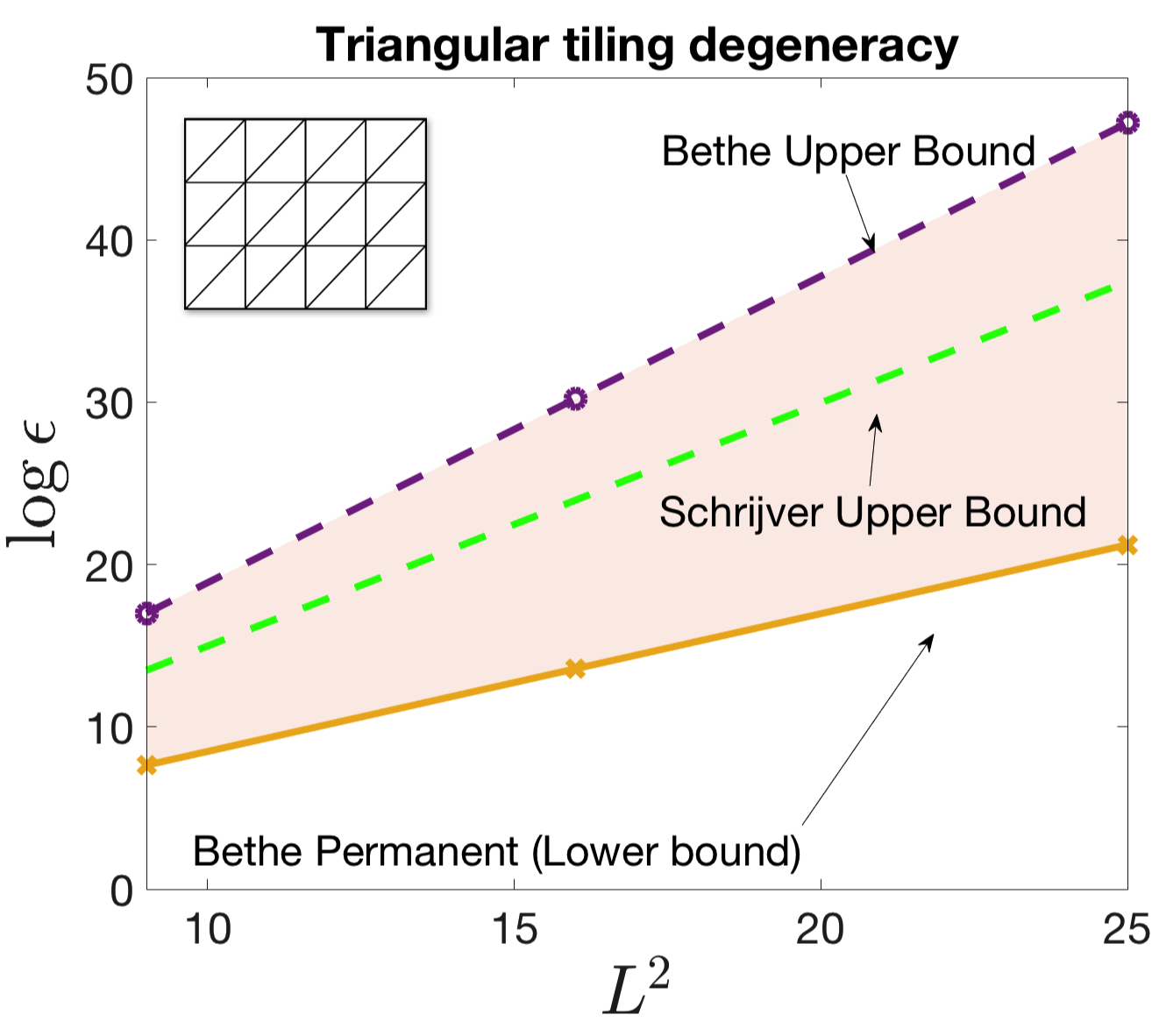}
    \caption{Numerical scaling of the degeneracy of the ground state via the Bethe Permanent for the square spin ice  and the triangular tiling (on the torus), of linear size $L$ (equal to the number of nodes). The lower bound (gold solid) corresponds to the Bethe Permanent $\text{Bperm}(A)$, while the upper bound (purple dashed) corresponds to $\sqrt{2^n} \text{Bperm}(A)$. A numerical fit shows that the Bethe permanent scales as $\epsilon_{SI}\approx \text{Bperm}(A_L)/(\prod (d_v/2)!)\approx (1.419[..])^{L^2}$, while Lieb's exact result is $\epsilon_{SI}^{Exact}=(\frac{8}{3\sqrt{3}})^{L^2}\approx (1.53[..])^{L^2}$. For the case of the triangular tiling, the scaling of the Bethe Permanent can be fit as $\epsilon_{TT}\approx (2.3396[..])^{L^2}$. In both figures, the shaded area is the bound on the (logarithm) of the degeneracy of the balanced configuration according to the Bethe Permanent bounds.}
    \label{fig:bethe}
\end{figure}

The numerical results are shown in Fig. \ref{fig:bethe} for the planar cases of the square lattice and the triangular tiling, which are two perfect Archimedean lattices \cite{harrison}. For the case of the square ice we have Lieb's exact result~\cite{lieb}  $\epsilon^{Exact}=(\frac{4}{3})^{3 L^2/2}$. For the case of the triangular tiling there is no exact solution. Pauling's argument \cite{pauling} does not apply, as it relies on the bipartiteness of the lattice. For the case of the triangular tiling, we find that $\frac{\ln \epsilon}{L^2} \geq 2.33[..]$, %\cn{(Ho cambiato a logaritmo naturale, controlla che vada bene). Considera anche che la degenerazione massima è sempre $2^{N_{spins}}$},
using the fact that the Bethe permanent is a lower bound.

\begin{figure}
    \centering
    \includegraphics[scale=0.18]{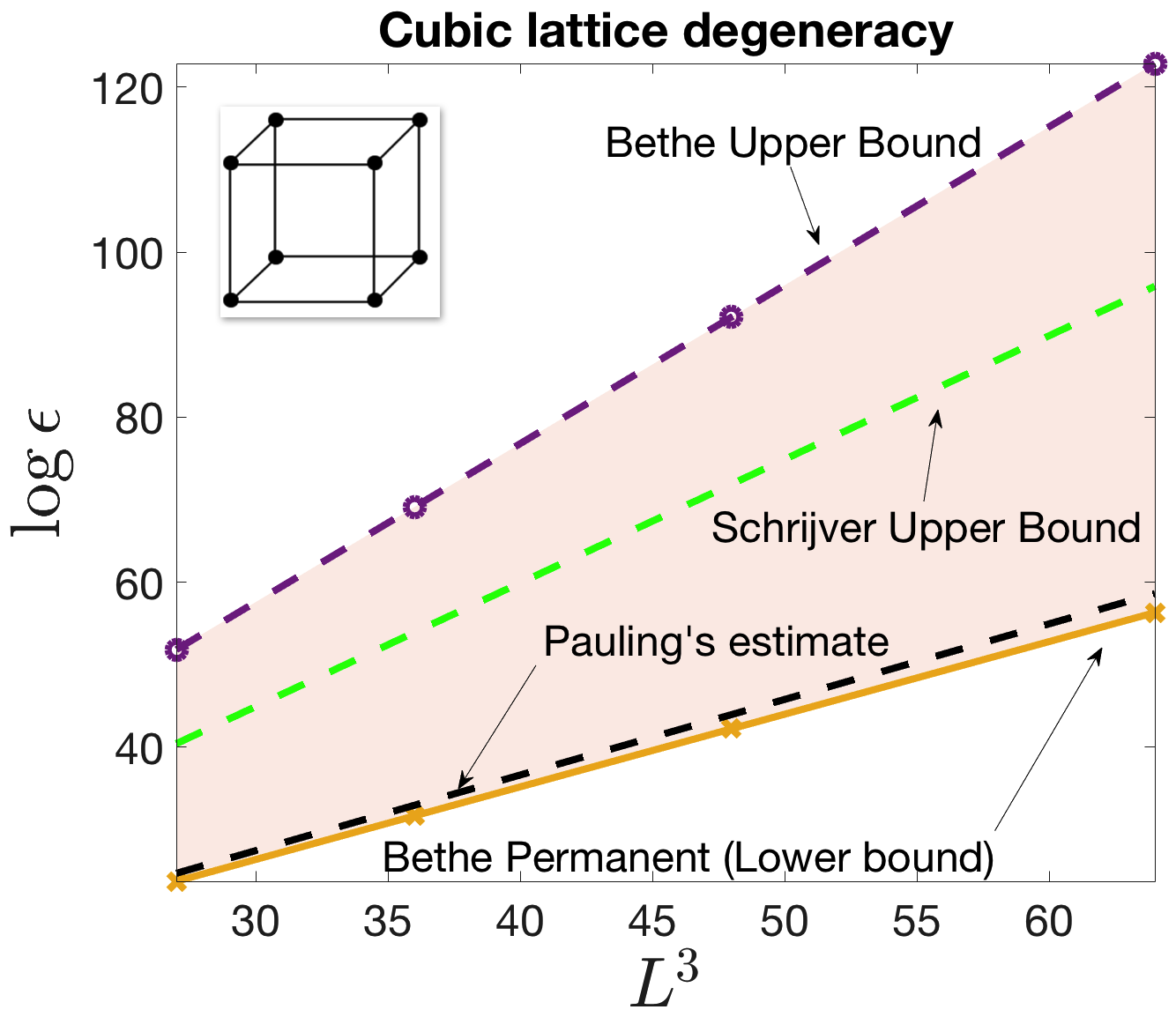}
    \caption{Numerical scaling of the degeneracy of the ground state via the Bethe Permanent for the cubic spin ice with a number of nodes $L^3$. The lower bound (gold solid) corresponds to the Bethe Permanent $\text{Bperm}(A)$, while the upper bound (purple dashed) corresponds to $\sqrt{2^n} \text{Bperm}(A)$. A numerical fit shows that the Bethe permanent scales as $\epsilon_{SI}\approx \text{Bperm}(A_L)/(\prod (d_v/2)!)\approx (2.41(0))^{L^3}$, while Pauling's estimate is $\epsilon_{\text{Pauling}}=(2.5)^{L^3}$. The lower bound obtained via the Bethe permanent is thus within $3\%$ of Pauling's estimate.}
    \label{fig:cubiclatt}
\end{figure}

For the cubic lattice (with toroidal boundary conditions), Pauling's calculation suggests that the entropy of the spin ice scales with the number of vertices $L^3$. % \cn{(N \`e il numero di vertici? Perché prima lo chiamavamo $L^2$ io preferisco $N$ ma allora bisogna essere consistenti (in seguito è chiamato invece $n$)} 
In fact, given a certain node, we have $2^6=64$ possible configurations, but only $20$ of them satisfy the ice rule. It follows that, from Pauling's argument, the spin ice ground state degeneracy should be

\begin{eqnarray}
\epsilon_{\text{Pauling}}=2^{3L^3} (\frac{20}{64})^{L^3}=(5/2)^{L^3}.
\end{eqnarray}
or $\ln \epsilon_{\text{Pauling}}=L^3 \ln \frac{5}{2}\approx 0.916(3)\cdot L^3$. 
Using the Bethe permanent (see Fig. \ref{fig:cubiclatt}), we observe that Pauling's estimate is not too far from the Bethe permanent lower bound, which gives $B\epsilon\approx 2.41^{L^3}$, but provides a certificate for a lower bound.

\begin{figure}[ht!]
    \centering
    \includegraphics[scale=0.18]{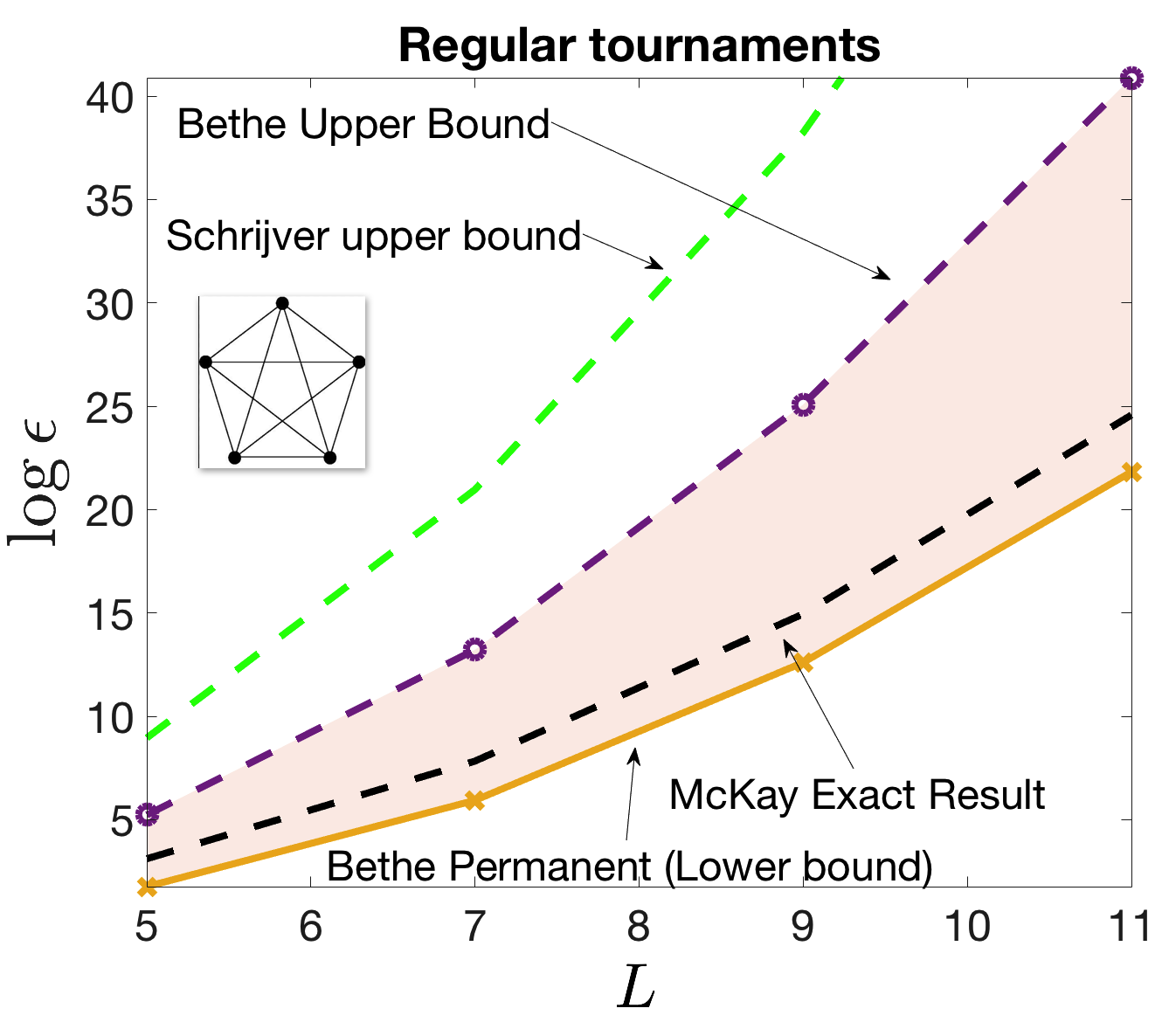}
    \caption{Number of Eulerian orientations for the complete graph $K_L$, also called regular tournaments. We provide a comparison between the upper and lower bounds given by the Bethe Permanent and the McKay exact calculation given in eqn. (\ref{eq:mckay}). We note that for this particular case, Schrijver upper bound is looser than the Bethe permanent upper bound. }
    \label{fig:regtourn}
\end{figure}

As a last application, we consider regular tournaments, which is the total number of Eulerian configurations for a complete graph with $L$ nodes, as in Fig. \ref{fig:regtourn}, or equivalently the degeneracy of the spin ice configurations on a complete graph. This number was calculated by McKay \cite{mckay} and is given by
\begin{eqnarray}
\epsilon=\left(\frac{2^{L+1}}{\pi L}\right)^{\frac{L-1}{2}} \frac{\sqrt{L}}{\sqrt{e}} \left[1+O(\frac{1}{\sqrt{L}})\right].
\label{eq:mckay}
\end{eqnarray}
It can be easily seen that also in this case the Bethe permanent provides a good estimate for the number of tournaments.

A summary of the results is provided in the Table \ref{tab:res}.

\section{Conclusions}

We have discussed some theoretical results for spin ice on arbitrary graphs. In the first part of the paper we have provided a graph theoretical mapping between spin ices and Ising models based on the directed incidence matrix of the Line graph. Specifically, we have shown something that was already known: while all spin ices can be mapped to Ising models, not all Ising models can be mapped to spin ices; but here we show a general a procedure to extract the original spin ice. We have also shown that all spin ices are degenerate (except the trivial case: the 1D Ising model). We have proved this result using the known gauge transformation for the Ising model. Another fact which we managed to prove in the first part of this paper, and that we found surprising, is that for planar spin ices, degeneracy is only at the vertex level. This implies that the effective Ising model, cycles between different vertices of the original spin ice are unfrustrated, as it can be seen via a gauge transformation to align these vertices. The same is not true for the interactions at the vertices, implying that the frustration scales with the number of vertices of the original spin ice. In the second part of the paper we have focused on techniques to bound the degeneracy of the ice manifold. The method is based on the number of Eulerian tours, as for every balanced graph (ice manifold state) there exists \textit{at least} one Eulerian trail. The advantage of using the Eulerian trails is that there are spectral techniques to count the number of Eulerian trails of a directed graph. We applied these techniques for even-degree and connected graphs, showing that these bounds can be extended to the case of odd-degree reducible graphs. To conclude, we have used an exact formulation for the degeneracy of spin ices configurations using a permanent identity. Given the fact that the permanent is $\#P$-complete quantity to compute, we have employed numerical methods based on the Bethe free energy. Such method provides numerical lower bounds to the permanent, and we have thus obtained lower bounds to the spin ice degeneracy for various spin ice regular lattices. The advantage of such procedure is that these certified lower bound estimates can be obtained relatively quickly and without using loop Monte Carlo techniques \cite{barkema,loop,giawei}, despite the latter giving a more precise estimate. As such, given two spin ice graphs, we can compare the lower bounds on these and provide a rough yet quick estimate of which will have a larger degenerate ground state.

It is worth mentioning that there are also other implementations using fractional belief propagation \cite{chertkov}. Such method depends on an extra parameter that (which is however matrix dependent), and is such that a particular value of this parameter coincides with  the exact result of the permanent. The key issue is that such parameter is not known a priori, but it is a promising venue for trying to estimate more precise values of the spin ice degeneracy. Such method will be considered in future papers. %\ms{values of the extra parameter give the bounds?) 

\begin{table}[]
    \centering
    \begin{tabular}{|c|c|c|c|c|}
    \hline
        \textit{Graph} & \textit{Exact} & \textit{Pauling} &  \textit{Bethe} & \textit{Maximum}\\
        \hline
        Square & 1.53[..] &  $\frac{3}{2}$ &  1.41 & $4$  \\
        Triangular &  &  & 2.33 & 8 \\
        Cubic &  & $\frac{5}{2}$ & 2.41 & 8 \\
       % Complete & x & x & x &\\
         \hline
    \end{tabular}
    \caption{Normalized ground state entropy $s=\frac{\ln \epsilon}{N_v}$. The Bethe permanent typically provides a certified lower bound given the matrix, and the scaling is extracted numerically. For the complete graph, tournaments do not scale simply exponentially with the number of vertices (see eqn. (\ref{eq:mckay})). The maximum is obtained via the relationship for regular graphs, $s= \ln 2^{\frac{d}{2}}$, with $N_v$ the total number of vertices.}
    \label{tab:res}
\end{table}

\begin{acknowledgements}
We thank Prof. A. Schrijver for some clarifications regarding the permanent formula for the eulerian orientations of eqn. (\ref{eq:perme}), and Prof. M. Chertkov for some clarifications regarding the Bethe Permanent. We also acknowledge the support of NNSA for the U.S. DoE at LANL under Contract No. DE-AC52-06NA25396. This work was carried out under the auspices of the U.S.
DoE through the Los Alamos National
Laboratory, operated by
Triad National Security, LLC
(Contract No. 892333218NCA000001). FC was also financed via DOE-LDRD grants PRD20190195.  M. Saccone acknowledges the Center for Nonlinear Studies for the fellowship which supports his research. 
\end{acknowledgements}

\end{document}